%% file: branching.tex
\documentclass{sig-alternate}

\usepackage[english]{babel}
\usepackage[latin1]{inputenc}
\usepackage{amssymb}
\usepackage{algorithm}
\usepackage{algpseudocode}
\usepackage{cite}
\usepackage{amsmath}
\usepackage{graphicx}
\usepackage[caption=false,font=footnotesize]{subfig}
\usepackage{url}
\usepackage{psfrag}
\usepackage[usenames]{color}
\usepackage{balance}
\usepackage{pstricks}
\usepackage{times}
\usepackage{ifpdf}
\usepackage[]{units}

\usepackage{blindtext}

\usepackage{etoolbox}
\makeatletter
\patchcmd{\maketitle}{\@copyrightspace}{}{}{}
\makeatother

\newcommand{\beq}{\begin{equation}}
\newcommand{\eeq}{\end{equation}}

\newcommand{\baq}{\begin{eqnarray}}
\newcommand{\eaq}{\end{eqnarray}}

\newcommand{\baqm}{\begin{eqnarray*}}
\newcommand{\eaqm}{\end{eqnarray*}}

\newcommand{\barr}{\begin{array}}
\newcommand{\earr}{\end{array}}

\newcommand{\bi}{\begin{itemize}}
\newcommand{\ei}{\end{itemize}}

\newcommand{\Exact}{\textsc{Exact}}
\newcommand{\Approx}{\textsc{Approximate}}

\newcommand{\balpha}{{\boldsymbol \alpha}}
\newcommand{\btheta}{{\boldsymbol \theta}}

\usepackage{array}

\makeatletter
\def\Ddots{\mathinner{\mkern1mu\raise\p@
\vbox{\kern7\p@\hbox{.}}\mkern2mu
\raise4\p@\hbox{.}\mkern2mu\raise7\p@\hbox{.}\mkern1mu}}
\makeatother

\newcommand{\techreport}[2]{#2}

\begin{document}

\title{Characterizing Branching Processes from Sampled Data}

\author{
	\fontsize{12}{12}\selectfont
 	\begin{tabular}{cc}
 		\multicolumn{2}{c}{Fabricio Murai$^{1}$, Bruno Ribeiro$^{1}$, Don
            Towsley$^{1}$, and Krista Gile$^{2}$} \tabularnewline
		&	\tabularnewline
		\fontsize{10}{10}\selectfont
		$^{1}$School of Computer Science& 
		\fontsize{10}{10}\selectfont
		$^{2}$Department of Mathematics and Statistics 		\tabularnewline
		\fontsize{10}{10}\selectfont
		University of Massachusetts & 
		\fontsize{10}{10}\selectfont
		University of Massachusetts \tabularnewline
		\fontsize{10}{10}\selectfont
		Amherst, MA, 01003 & 
		\fontsize{10}{10}\selectfont
		Amherst, MA, 01003		\tabularnewline
		\fontsize{10}{10}\selectfont
		\{fabricio, ribeiro, towsley\}@cs.umass.edu  & 
		\fontsize{10}{10}\selectfont
		gile@math.umass.edu
 	\end{tabular}
}

\maketitle

\input{abstract}

%


\input{introduction}
\input{model}
\input{estimation}

\balance

\input{results}
\input{conclusions}


\bibliographystyle{plain}
\bibliography{branching}

\techreport{\input{appendix}}{}

\end{document}

%% file: abstract.tex
\begin{abstract}
Branching processes model the evolution of populations of agents that randomly generate offsprings.
These processes, more patently Galton-Watson processes, are widely used to model biological, social, cognitive, and technological phenomena, such as the diffusion of ideas, knowledge, chain letters, viruses, and the evolution of humans through their Y-chromosome DNA or mitochondrial RNA.
A practical challenge of modeling real phenomena using a Galton-Watson process is the offspring distribution, which must be measured from the population.
In most cases, however, directly measuring the offspring distribution is unrealistic due to lack of resources or the death of agents.
So far, researchers have relied on informed guesses to guide their choice of offspring distribution.
In this work we propose two methods to estimate the offspring distribution from real sampled data.
Using a small sampled fraction of the agents and instrumented with the identity
of the ancestors of the sampled agents, we show that accurate offspring
distribution estimates can be obtained by sampling as little as 14\% of the population.

\end{abstract}

%% file: introduction.tex

\section{Introduction} \label{sec:intro}

Branching processes, more markedly Galton-Watson (GW) processes, have been used to
model a variety of phenomena, ranging from human Y-chromosome DNA and
mitochondrial RNA evolution~\cite{neves2006}, to epidemics on complex
networks~\cite{vespignani}, to block dissemination in peer-to-peer
networks~\cite{veciana2004}. 
The GW process can be represented as a growing tree, where agents are nodes connected to their offspring by edges.
The number of offspring is a random variable associated with a distribution function.
An example of a GW branching process is a family tree considering either only the females or only the males in the family (which represent the transmission of mitochondrial RNA or Y-chromosome DNA, respectively). 
A GW process is completely characterized by its offspring distribution.
A practical challenge when modeling real world systems from a GW process is knowing the offspring distribution of the process, which must be measured from the population.

In most applications, however, directly measuring the offspring distribution is unrealistic due to the lack of resources or the inaccessibility of agents (e.g.\ death).
It is not reasonable to assume that one can collect genetic material from the entire human population or that
in the branching process of chain letter signatures (see Chierichetti et al.~\cite{Chierichetti11} for further details), one may collect all possible branches of the chain letter created by forwarding the letter.
So far, researchers have relied on informed guesses to guide their choice of
offspring distribution.
%


In this work we propose a collection of methods to estimate the offspring distribution from real sampled data.
Our goal is to accurately estimate the offspring distribution by sampling and collecting ancestors ids of a small fraction of the agents.
We study the case where a sampled agent reveals the identity of its ancestors and 
the trees are generated in the supercritical regime (i.e., average offspring > 1)
when the maximum offspring and the maximum tree height are upperbounded by a (possibly large) constant.
We show that accurate offspring distribution estimates can be obtained by
sampling as little as 14\% of the population.

A related problem is characterizing graphs using traceroute sampling.
Traceroute sampling from a single source can be thought as sampling a tree where
nodes have different offspring (degree) distributions depending on their
position with respect to the source. 
This is an important well known hard problem~\cite{Lakhina03,Achlioptas09} and it remains open to date. 
Our results have the added benefit of shedding some light also into the traceroute problem.

The outline of this work is as follows.
Section~\ref{sec:model} describes
the network and sampling models. 
In Section~\ref{sec:estimation} we first show how to estimate the offspring distribution
through exact inference, showing it does not scale. 
We then propose an MCMC method of performing approximate inference that works for
small and medium sized trees (up to 2,000 nodes).
In Section~\ref{sec:results} we evaluate both methods using a set of 900 syntethic datasets,
comprising small and medium trees.
For small trees, 
exact inference yielded accurate estimates and outperforms the
approximate estimator. On the other hand, approximate inference can handle
larger trees, while obtaining significant improvement over more na\"ive approaches.
Finally Section~\ref{sec:conclusions} presents our conclusions and future work.

%% file: model.tex
\section{Model} \label{sec:model}

We assume that the underlying tree comes from a Galton-Watson (GW) process.
The GW process models the growth of a population of
individuals that evolves in discrete-time ($n=0,1,2,\dots$) as follows.
The population starts with one individual at the $0$-th generation ($n=0$).
Each individual $i$ at the $n$-th generation
produces a random number of individuals at the $(n+1)$st generation, called
offspring. The offspring counts of all individuals are assumed to be i.i.d.
random variables. 
An instance of the GW process is therefore described by a
sequence of integers which denote the number of individuals at each generation.

Formally, the GW process is a discrete-time Markov Chain\newline $\{X_n\}_{n=1}^L$, where $L$ is the number of generations, given by the
following recursion
\begin{equation*}
 X_{n+1} = \sum_{i=1}^{X_{n}}Y_{i}^{(n)} \, ,
\end{equation*}
with $X_0 = 1$, where the $Y_{i}^{(n)}\geq 0$ are i.i.d.\ random variables
with distribution $\btheta=(\theta_0,\dots,\theta_W)$, $\forall i, n \geq1$, where $W$ is the maximum number of offspring of an agent.
The GW process can be seen as a  generative process of a tree $G=(V,E)$,
where $X_n$ is the number of nodes at the $n$th generation
and $Y_i^{(n)}$ is the offspring count of the $i$th node at the $n$th generation.
For simplicity, we assume that $\theta_0 = 0$ and that the number of generations is
fixed, so that all tree leaves sit exactly at generation $L$. 
Our results, however, can be easily adapted to the case where  $\theta_0 > 0$ and the leaves have different levels.
But the above assumptions lead to a simpler model in the sense that we can have average offspring greater than one without
worrying about infinite trees.
%

\begin{figure}[!t]
\centerline{
\subfloat[Original]{\includegraphics[width
=0.8in]{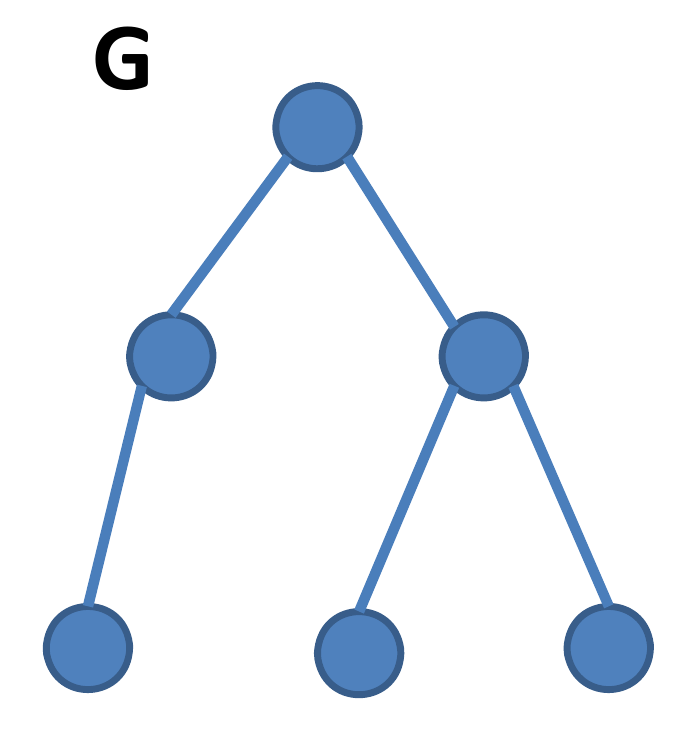}
\label{fig:network}}
\hspace{1.0cm}
\subfloat[Examples of samples]{\includegraphics[width
=1.4in]{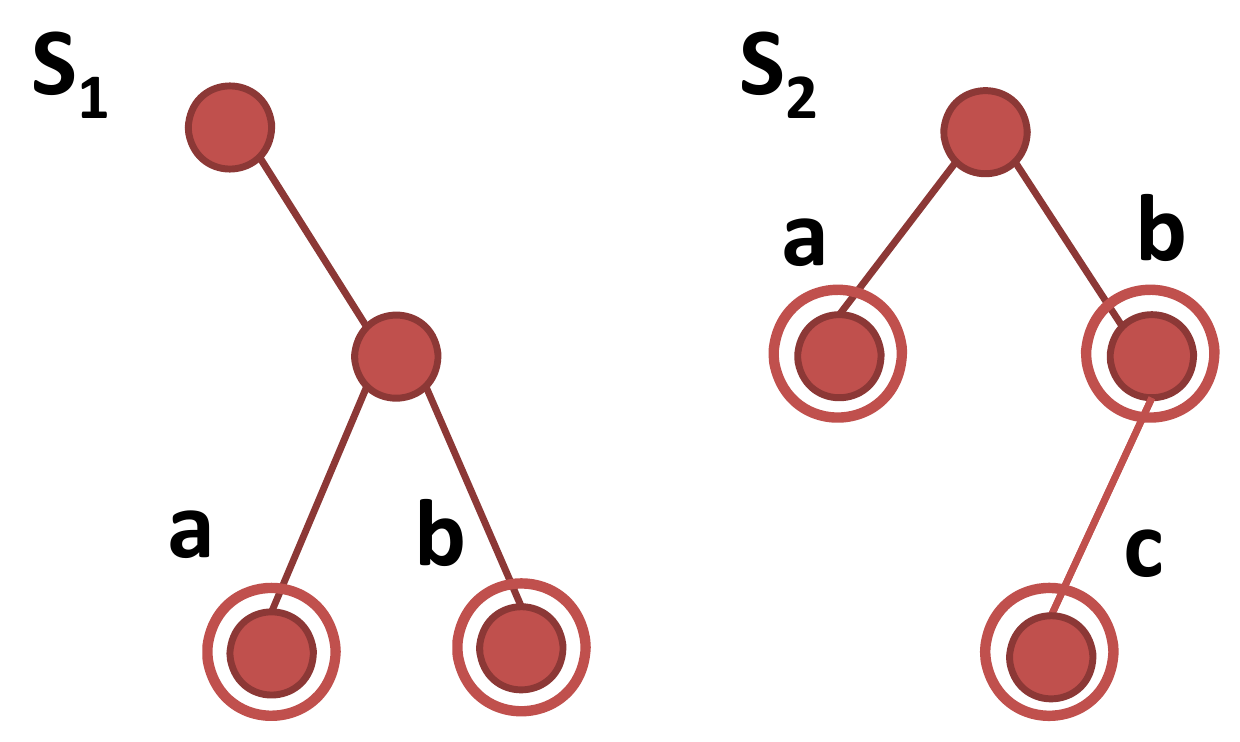}
\label{fig:samples}}
}
\caption{(a) Branching process tree. (b) Samples with 2 and 3 targets,
respectively.}
\label{fig:model}
\end{figure}

Since the numbers of offspring are mutually independent,
the probability of a given tree $G$ is
\begin{equation}
P(G|\btheta) = \prod_{j=1}^W \theta_{j}^{c_j} \, ,
\end{equation}
where $c_j=\sum_{i,n} \mathbf{1}\{Y_i^{(n)}=j\}$ is the number of nodes with
offspring count $j$.
Fig.~\ref{fig:network} depicts an example of
tree generated from $\btheta=(0.3,0.6,0.1)$ with $L=3$.
In this case, $P(G|\btheta) = 0.3^1\cdot0.6^2 = 0.108$.


\subsection*{Sampling Model}

A node is said to be {\it observed} when the sampling process
explicitly reveals its presence in the original graph (e.g. node look up is performed
or node spontaneously advertise its presence).
The {\it observed path}, however, consists of the observed node and its path to the root.
A {\it sample} is a set of observed paths.

Let $V^\prime \subset V$ be a set of randomly observed nodes of the unlabeled
graph $G$.
Let $S$ be the sampled tree formed by the union of the paths from all nodes $v \in V^\prime$ to the root of $G$. For instance,
Fig.~\ref{fig:samples} shows sampled trees $S_1$ formed by $V^\prime=\{a,b\}$
and $S_2$ formed by $V^\prime = \{a,b,c\}$.
We assume that nodes in $V^\prime$ are sampled from $V$ with probability $p$.



We now show how to compute $P(S|G)$, assuming that 
that $V^\prime$ is known, i.e., we know which nodes in $S$ are observed.
However, it is easy to modify the following analysis to the cases where
(1) we only know $|V^\prime|$ or (2) we know the topology of $S$,
but not which or how many nodes are observed. Let $C_{G,S}$ be the number of ways 
in which $S$ can be mapped onto
$G$. Clearly, $C_{G,S} = 0$ if $S$ is not a subgraph of $G$.
Conditioning on a given mapping, we must have exactly $|V^\prime|$
nodes chosen as targets and $|V\setminus V^\prime|$ not chosen as so.
Therefore,
\begin{equation}\label{eq:probSgivenG}
P(S|G)= C_{G,S}\ p^{|V^\prime|}(1-p)^{|V\setminus V^\prime|}.
\end{equation}

Computing $C_{G,S}$ can be done recursively by first computing $c_{ij}$, the
number of ways the $i$-th subtree connected to the root of $S$ can be mapped to
$j$-th subtree connected to the root of $G$, for all $i, j$. Now consider the matrix
$\mathbf{C} =
[c_{ij}]_{n \times m}$. If we define the operator
\beq
|\mathbf{C}_{n\times m}|= \begin{cases}
\sum_{j=1}^m c_{1j} |\mathbf{C}_{1j}|, & n \geq 1\\
\sum_{j=1}^m c_{1j}, & n = 1, \end{cases}
\eeq
where $\mathbf{C}_{1j}$ is $\mathbf{C}$ after removal of the 1st row and $j$th
column, then we can show that $C_{G,S} = |\mathbf{C}|$. Consider the simple case
of $G$ and $S_2$ shown in Fig.~\ref{fig:samples}. Here we have
$\mathbf{C}=\left[\begin{array}{cc}
1 & 1\\
2 & 1\end{array}\right]$
and hence, $|\mathbf{C}| = 1 \cdot 1 + 1 \cdot 2 = 3$. We can visually check
that this is indeed the number of ways to map $S_2$ onto $G$. Therefore, $P(S_2|G) =
3 p^3 (1-p)^3$.

Inference on the structure of the tree $G$ from the partial observation $S$
is possible because we can compute $P(S|G')$ for any $G'$. This, in turn, allows
us to do inference on the offspring distribution by considering how likely $G'$ is to be
generated from $\btheta$ by using $P(G'|\btheta)$ and weighting by how likely
$S$ is to be sampled given $G'$. In the next section we propose two estimation
methods based on this idea.

%% file: estimation.tex
\section{Estimators} \label{sec:estimation}

We consider the problem of estimating the offspring distribution $\btheta$ of the
GW process that generates a tree $G$ given a sample $S$
consisting of the union of random observed paths when nodes are observed with probability $p$.

Two approaches to this problem based on Maximum Likelihood Estimation are proposed in this paper.
While the former consists of the exact computation of
the likelihood function $P(S|\btheta)$, the latter approximates this function
via Metropolis-Hastings with importance sampling.

\subsection{Exact inference}

\begin{figure}[!t]
\centering
\includegraphics[width=2in]{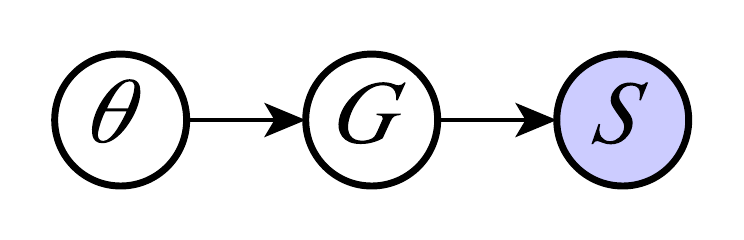}
\caption{Graphical model representing network generation and sampling. White
    nodes are unobservable and shaded node is observable.}
\label{fig:graphModel}
\end{figure}

The graphical model in Fig.~\ref{fig:graphModel} depicts the statistical relationship between $S$, $\btheta$ and $G$. 
The
shaded node, $S$, is the only observable variable, while the white
nodes, $\btheta$ and $G$ are unobservable. This figure shows that to
find the relationship between $S$ and $\btheta$, we have to sum over the
variable $G$, i.e., over all possible trees given the number of generations $L$ and
the maximum degree $W$. Let $\mathcal{G}_{L,W}$ be the set of all possible trees
given $L$ and $W$. It follows that
\begin{eqnarray}
P(S|\btheta) & = & \sum_{G \in \mathcal{G}_{L,W}}P(S,G|\btheta)\nonumber\\
 & = & \sum_{G \in \mathcal{G}_{L,W}}P(S|G,\btheta)P(G|\btheta)\nonumber\\
 & = & \sum_{G \in \mathcal{G}_{L,W}}P(S|G)P(G|\btheta), \label{eq:lik}\end{eqnarray}
where from line 2 to line 3 we use the fact that $S$ is conditionally
independent on $\btheta$ given $G$ (see Fig.~\ref{fig:graphModel}).
However, $|\mathcal{G}_{L,W}|$ grows exponentially both in
$L$ and $W$,
which limits this approach to very small trees.
In fact, we can show that
\[
|\mathcal{G}_{L,W}|=\begin{cases}
W, & L=1\\
\sum_{i=1}^{W}|\mathcal{G}_{L-1,W}|^{i}>|\mathcal{G}_{L-1,W}|^{W}, & L>1.\end{cases}\]
Solving the recursion yields
$\log^{(L-1)} |\mathcal{G}_{L,W}| = O(W)$,
where $\log^{(.)}$ is the repeated logarithm. Note however that 
isomorphic trees are being counted more than once. Therefore, we can reduce the
computational cost by counting only non-isomorphic trees (appropriately weighted by their
multiplicity).

\begin{table}
\renewcommand{\arraystretch}{1.3}
\label{tab:spaceGrowth}
\centering
\begin{tabular}{c|c|c|c|c|c}
\hline 
$L$ & 1 & 2 & 3 & 4 & 5\tabularnewline
\hline
\hline 
$|\mathcal{G}_{L,3}| \approx$ & 3 & 39 & $6\times10^{4}$ & $2.3\times10^{14}$ & %
$1.2\times10^{43}$\tabularnewline
\hline 
$|\mathcal{G}_{L,3}^{\textrm{non-iso}}| \approx$ & 3 & 19 & $1.5\times10^{3}$ & $6.1\times10^{8}$ & %
$3.8\times10^{25}$\tabularnewline
\hline
\end{tabular}
\caption{Growth of the space of trees as a function of $\mathbf{L}$, for $\mathbf{W=3}$.}
\end{table}

Let $\mathcal{G}_{L,W}^{\textrm{non-iso}}$ be the maximal set of non-isomorphic
trees of $\mathcal{G}_{L,W}$.
It is possible to show that 
\[
|\mathcal{G}_{L,W}^{\textrm{non-iso}}|=\begin{cases}
W, & L=1\\
\frac{(W+1)\binom{W+|\mathcal{G}_{L-1,W}^{\textrm{non-iso}}|}{W+1}}{|\mathcal{G}_{L-1,W}^{\textrm{non-iso}}|}-1, & L>1.\end{cases}\]
Table~\ref{tab:spaceGrowth} illustrates some values of $|\mathcal{G}_{L,W}|$ and $|\mathcal{G}_{L,W}^{\textrm{non-iso}}|$
for $W=3$ and $L=1,\dots,5$.
As we can see, counting only non-isomorphic trees reduces significantly
the state space, but it is still not feasible to compute eq.~\eqref{eq:lik}
except for rather small numbers such as $W=3$ and $L=4$.
Nevertheless,
we utilize this approach to perform inference more efficiently.
In the following, we
explain how to enumerate trees in $\mathcal{G}_{L,W}^{\textrm{non-iso}}$ and how
to compute their multiplicities.

\subsubsection*{Counting only non-isomorphic trees}

A straightforward way to enumerate all trees
in $\mathcal{G}_{L,W}^{\textrm{non-iso}}$ is: (1) to enumerate non-isomorphic trees
in $\mathcal{G}_{L-1,W}^{\textrm{non-iso}}$ and assign a numeric
id to each of them; and (2) construct trees in
$\mathcal{G}_{L,W}^{\textrm{non-iso}}$ by attaching to a root node
trees from $\mathcal{G}_{L-1,W}^{\textrm{non-iso}}$ where ids are
in non-increasing order. Note that two trees are isomorphic in this
construction if the sets of ids of the subtrees connected to the root
node are permutations of each other, which cannot occur due to the
ordering. 

In what follows we compute the probability that sample $S$ is observed given
the offspring distribution $\btheta$ through the enumeration of non-isomorphic
trees. Let $m_{i}^{(L)}$ denote the multiplicity of the $i$-th tree, say $G_{i}$,
in the labeled space $\mathcal{G}_{L,W}^{\textrm{non-iso}}$. Eq.~\eqref{eq:lik} is equivalent to 
\begin{equation}
P(S|\mathbf{\theta})=\sum_{G_{i}\in\mathcal{G}_{L,W}^{\textrm{non-iso}}}m_{i}^{(L)}P(S|G_{i})P(G_{i}|\mathbf{\theta}).\label{eq:lik-noniso}\end{equation}
The multiplicity $m_{i}^{(L)}$ can be calculated from the ids of subtrees
directly connected to the root node in $G_i$ and their multiplicities. More
precisely, $m_{i}^{(L)}$ is simply the number of permutations of the
ids multiplied by the product of the multiplicities of each subtree. For instance,
if there are $j$ subtrees connected to the root with distinct ids $(1),\dots,(j)$, then
$m_{i}^{(L)}=j!\times\prod_{k=1}^{j}m_{(k)}^{(L-1)}$.
In the general case, where ids can appear more than once, we have
\[
m_{i}^{(L)}=\frac{j!\times\prod_{k=1}^{j}m_{(k)}^{(L-1)}}
{\prod_{id=1}^{|\mathcal{G}_{L-1,W}^{\textrm{non-iso}}|}\left(\sum_{k=1}^{j}\mathbf{1}\{id=k\}\right)!}.\]


The first
estimator we propose is
\beq
\hat{\btheta}_{\textrm{Exact}} = \arg \max_\btheta P(S|\btheta),\label{eq:exact}
\eeq
where $P(S|\btheta)$ is computed as in \eqref{eq:lik-noniso}.

\subsubsection*{Maximum Likelihood Estimation}

After obtaining a sample, we write the summation in Eq. (\ref{eq:lik-noniso})
as a function of $\btheta$. Unfortunately, this likelihood
function is a sum of a potentially enormous number of terms and using
the log-likelihood is not helpful in this case. We apply several tricks
to solve this optimization task.

One simple trick to reduce the number of terms consists of grouping
together trees that have the same configuration in terms of offspring
counts, i.e., that account for the same $P(G|\mathbf{\theta})$. Note
that there are many such trees even when considering non-isomorphic
trees only, although they correspond to different values of $P(S|G)$.

Also note that this is a constrained maximization problem. Since $\btheta$
is a probability distribution, $0\leq\theta_{i}\leq1$ for $i\in\{1,\dots,W\}$
and $\sum_{i=1}^{W}\theta_{i}=1$. We can turn it into a non-constrained
maximization problem by replacing $\theta_{i}=\frac{e^{\alpha_{i}}}{Z}$
for $i\in\{1,\dots,W\}$ where $Z=\sum_{i=1}^{W}e^{\alpha_{i}}$,
setting $\alpha_{W}=1$ (for regularization purposes) and then maximizing
w.r.t.\ $\balpha$. Note that $\mathbf{\alpha}_{i}$
can now assume any value in $\mathbb{R}$ for $i\in\{1,\dots,W-1\}$.
Nevertheless, one must be careful when using this parameter transformation
since the products of the exponentials can quickly lead to overflows.
Therefore, we use log representation and the {\it logsumexp} trick.

After this transformation, the maximization problem becomes
\[
\max_{\alpha}l(\alpha)=\sum_{j}c_{j}\frac{e^{\sum_{i=1}^{W}x_{ji}\alpha_{i}}}{Z^{y_{j}}}\]
where $c_{j}$ is the sum of the coefficients of the terms corresponding to the
same $j$-th configuration of $P(G|\theta)$, $x_{ji}$ is the number of
nodes with offspring $i$ in the $j$-th configuration and $y_{j}=\sum_{i=1}^{W}x_{ji}$.
In order to compute the likelihood function and its gradient
more efficiently, we express them in matrix notation as
\[
l(\alpha)=\mathbf{c}^{T}\cdot(\exp(\mathbf{X}\balpha)/Z^{\mathbf{y}})\]
\[
\nabla l(\alpha)=\exp(\mathbf{X}\balpha)/Z^{\mathbf{y}}-\exp(\mathbf{X}\balpha)/Z^{\mathbf{y}+\mathbf{1}}\]
where $\mathbf{c}=[c_{j}]$, $\mathbf{X}=[x_{ji}]$, $\balpha=[\alpha_i]$, $\mathbf{y}=[y_{j}]$,
$Z^{\mathbf{y}}=[Z^{y_{j}}]$, the ``$/$'' symbol corresponds to
division of two vectors element-wise and $\mathbf{1}$ is a column
vector with all entries equal to 1.

The maximization then goes as follows. We sample $10,000$ points uniformly
from $\mathbb{R}^{W-1}$. The one with the maximum value of $l(.)$
will be $\balpha^{(0)}$, the starting point to be used with
the BFGS\footnote{We use R implementation in package \textbf{stats}.}
(limited to 100 iterations, relative convergence tolerance of $10^{-8}$,
step size $10^{-3}$). The estimate $\hat{\btheta}$ can be obtained
from $\hat{\balpha}$ by exponentiating and then normalizing
the latter.

\subsection{Approximate inference with MCMC}\label{sec:approx}

The previous approach only applies to small problems due to the enormous number
of terms in the summation \eqref{eq:lik-noniso}.
To solve larger problems, we approximate eq.~\eqref{eq:lik} using
MCMC.

Let $h=P(S|G)$ and $f(G)=P(G|\btheta)$. Since $f(G)$ defines a probability
distribution on the space $\mathcal{G}_{L,W}$, it follows that
\begin{equation}
P(S|\btheta) = \sum_{G} h f(G) = E_f[h].
\end{equation}
where $E_f[.]$ denotes expectation w.r.t. distribution $f$.

Monte Carlo simulation approximates expectations (integrals, more generally) by
sampling from a desired distribution $f$~\cite{Beerli99}. The problem here is that we cannot sample from $f$
because we don't know $\btheta$. However, we can sample from some other
distribution $g$ and compensate for the fact that in $g$ some trees are more
(or less) likely to appear than in $f$ by using importance sampling. More
precisely,
\beq \label{eq:expect}
P(S|\btheta) = \sum_G h f(G) = \sum_G h \frac{f(G)}{g(G)}g(G) =
E_g\left[h\frac{f(G)}{g(G)}\right].
\eeq

Recall from Section~\ref{sec:model} that we can generate trees using the GW
process from a given offspring distribution $\btheta_0$. Hence we can set
\beq \label{eq:g}
g(G) = \frac{1}{\mathcal{Z}} P(S|G)P(G|\btheta_0)\, ,
\eeq
where $\mathcal{Z}$ is a normalizing constant\footnote{We could have set
$g(G)=P(G|\btheta_0)$ instead, but our approach restrict us to
generating trees that are consistent with the sample and thus, is more
efficient.}.
Substituting eq.~\eqref{eq:g} into \eqref{eq:expect} yields
\begin{equation*}
P(S|\btheta) =
E_g\left[\frac{P(S|G)P(G|\btheta)}{\frac{1}{\mathcal{Z}}P(S|G)P(G|\btheta_0)}\right]
\approx \frac{\mathcal{Z}}{m} \sum_{i=1}^m
\frac{P(G_i|\btheta)}{P(G_i|\btheta_0)},
\end{equation*}
where $G_i \sim g(G)$. Note that $\frac{\mathcal{Z}}{m}$ is not a function of $\btheta$
and do not need to be considered when maximizing $\btheta$. Therefore, the
second estimator we propose is
\beq
\hat{\btheta}_{\textrm{Approximate}} = \arg \max_\btheta \sum_{i=1}^m
\frac{P(G_i|\btheta)}{P(G_i|\btheta_0)}.
\eeq

In order to draw $G_i \sim g(G)$, we use the Metropolis-Hastings algorithm where
each state $X_j$ of the Markov Chain is a tree. We
start the chain in a state $X_0$ consistent with $S$, in particular, we set
$X_0=S$. The transition kernel $X_i \rightarrow X_{i+1}$ we use is shown in
Algorithm~\ref{alg:transition}.
\begin{algorithm}
\caption{Transition Kernel($X_i,X_{i+1}$)}
\label{alg:transition}
\begin{algorithmic}[0]
\State $v \leftarrow $ internal node selected uniformly at random from $X_i$
\State $d_v \leftarrow $ degree($v$)
\If{$d_v = 1$} \State $action \leftarrow add$

\ElsIf{$d_v = W$}
        \State  $action \leftarrow remove$
    
\Else \Comment{$1<d_v<W$}
    \If{$ U(0,1) < 0.5$} \Comment{$U(0,1)$ is the uniform dist.}
        \State $action \leftarrow add$
    
    \Else
        \State $action \leftarrow remove$
    \EndIf
\EndIf

\If{$action = add$}
    \State $T_v \leftarrow$ GaltonWatson($\btheta_0,L-l$)
    \State $v.child[d_v+1] \leftarrow T_v$ \Comment{adds new branch}
    \State $d_v \leftarrow d_v+1$

\ElsIf{$action = remove$}
    \State shuffle($v.child$) \Comment{shuffle children}
    \State $v.child[d_v] \leftarrow $ nil\Comment{removes ``right-most'' branch}
    \State $d_v \leftarrow d_v-1$
\EndIf
\end{algorithmic}
\end{algorithm}
The new tree $X_{i+1}$ is accepted with probability
\beq
r = \min\left(1,\frac{P(S|X_{i+1})P(X_{i+1}|\btheta_0) q(X_{i+1}\rightarrow
            X_i)}{P(S|X_i)P(X_{i}|\btheta_0)q(X_{i}\rightarrow X_{i+1})}\right).
\eeq
where $q(X_i \rightarrow X_j)$ is the probability that the transition kernel
proposes transition $X_i \rightarrow X_j$. It is easy to include the calculation
of $q(X_i \rightarrow X_{i+1})$ and $q(X_{i+1} \rightarrow X_{i})$
in the transition kernel implementation. In particular, let $N_i$ and $L_i$
denote the number of nodes and leaves in $X_i$, respectively. Hence,
if $action = add$,
\begin{eqnarray*}
q(X_{i}\rightarrow X_{i+1})&=&\frac{0.5^{\mathbf{1}\{d_v>1\}}\times
P(T_{v}|\theta_{0})}{N_{i}-L_{i}-1},\\
q(X_{i+1}\rightarrow
        X_{i})&=&\frac{0.5^{\mathbf{1}\{d_{v}+1<W\}} (d_v+1)^{-1}}{N_{i+1}-L_{i+1}-1},
\end{eqnarray*}
otherwise,
\begin{eqnarray*}
q(X_{i}\rightarrow X_{i+1})&=&\frac{0.5^{\mathbf{1}\{d_{v}<W\}}d_v^{-1}}{N_{i}-L_{i}-1},\\
q(X_{i+1}\rightarrow
        X_{i})&=&\frac{0.5^{\mathbf{1}\{d_{v}-1>1\}}\times
P(T_{v}|\theta_{0})}{N_{i+1}-L_{i+1}-1},
\end{eqnarray*}
where $0.5^{\mathbf{1}\{d_v>1\}}$ accounts for the fact that if $v$ has degree
$> 1$, action {\it add} is chosen with probability $0.5$, but when $d_v = 1$,
{\it add} is always chosen. The case for {\it remove} is similar\footnote{All
calculations should be performed in log space.}.


\subsubsection*{Maximum Likelihood Estimation}
After obtaining roughly independent samples $G_i \sim g(G)$, we write the summation in the RHS of
eq.~\eqref{eq:expect} and perform maximization as in the case of exact
inference.

%% file: results.tex
\section{Experiments and Results} \label{sec:results}

We first describe the experiments used to assess the
performance of the two estimation
methods, henceforth referred to as \Exact\ and \Approx, respectively.
We then compare methods w.r.t.\
the KL-divergence of the estimated distribution from $\btheta$.
In addition, we show some results in detail to illustrate the Mean Squared Error
(MSE) per distribution parameter and how performance increases with the
sampling probability. In general, \Exact\ performs best but is only feasible
for small datasets. Nevertheless, \Approx\ exhibits comparable performance and
can cope with larger datasets (up to 2,000 nodes).

\subsection{Experiments description}

Based on the size of $\mathcal{G}_{L,W}$, we define two
classes of estimation problems: small and medium size problems.
For medium size ones, we would like to compare the methods' performance
for short and long tail offspring distributions, hereby represented by truncated\footnote{Here truncated means
that we took the original probability mass function for values between $1$ and
$W$ and normalized by their sum, while setting the probability mass of other values to zero.}
Poisson and Zipf distributions, respectively. Parameters of these
distributions were chosen so that their average is $\bar{d}$.

In what concerns
the sampling process, we choose three sampling probabilities representing low,
medium and high sampling rates for each class.
The set of values of $p$ has to be different for each class for two reasons.
The practical reason is that as the tree size grows, the cost to sample it grows
linearly on $p$ and we may be limited by a budget. The second reason is that,
if there is no such constraint, while values of $p$ such as $0.5$ are reasonable for small problems,
they will likely reveal all nodes from the top levels for large problems. Hence,
taking the empirical distribution from the first levels per se would be an
accurate estimator.
Inside each class,
consider the following distributions and sampling probabilities:
\begin{enumerate}
  \setlength{\itemsep}{1pt}
  \setlength{\parskip}{0pt}
  \setlength{\parsep}{0pt}
\item Small size: $W=3, L=3, \bar{d} = 2.1$
\begin{itemize}
\item $\btheta^{(1)}=(0.2,0.5,0.3)$
\item $p \in \{0.1,0.2,0.5\}$
\end{itemize}
\item Medium size: $W=10, L=5, \bar{d}=3.15$
\begin{itemize}
\item $\btheta^{(2)} \sim$ truncated Poisson($\lambda=3$)
\item $\btheta^{(3)} \sim$ Zipf($\alpha=1.132$, $N=10$)
\item $ p \in \{0.5,1.0,5.0\}\times10^{-2}$
\end{itemize}
\end{enumerate}
Average tree sizes per class are $\approx 17$ and $\approx 454$, respectively.

In order to test the inference methods, we build a set of estimation problems as follows.
For each distribution
$\btheta^{(i)},\,i=1,\dots,3$, we generate 10 trees $t_{ij},\,j=1,\dots,10$ from a
GW process with height $L+1$ (30 trees in total). Next, for each of
the 90 pairs $(t_{ij},p_{ik}),\,k=1,2,3$, we generate 10 samples
$s_{ijkl},\,l=1,\dots,10$ (900 samples in total).


We assume each sample $s_{ijkl}$ constitutes a separate estimation problem 
(also referred to as {\it dataset} to avoid confusion with MCMC samples). This can be
interpreted as if we had one tree (originated from the GW process), and a single
opportunity to sample it.
No other samples can be obtained from the same tree, nor other trees
are available for sampling. Ideally, we would like to try both
methods with each problem, but \Exact\ is only feasible for small problems.
Before presenting the results, we briefly discuss implementation issues related
to \Approx.


\subsection{Implementation issues of APPROXIMATE}

The main difficulty in the \Approx\ method is knowing when to stop the approximation as, without knowing the true distribution, we need a mechanism that tells us how close we are to the steady state distribution of the Markov chain.

Recall that we use the Metropolis Hastings (MH) algorithm to sample graphs from $g(G)$ (see
Eq.~\eqref{eq:expect}). As with any MCMC method, three questions must be addressed: (1) How long
should the burn-in period be? (2) What should the thinning ratio be? (3) What is
the minimum number of uncorrelated samples that we need?
We use the Raftery-Lewis (RL) Diagnostic \cite{raftery95} to address these issues\footnote{We use the R
implementation in package {\bf coda}.}.

The RL Diagnostic attempts to determine
the necessary conditions to estimate a
quantile $q$ of the measure of interest, within a tolerance $r$ with probability
$s$. We take the likelihood of the MH samples as the measure of interest.
The diagnostic was then applied individually to each dataset with default
parameters ($q=0.025$, $r=0.005$ and $s=0.95$).
Results concerning the burn-in period and thinning ratio are subsumed by the
required number of samples and hence will be ommited.
The minimum number of MCMC samples for small datasets was less than 50,000 graphs and
for medium datasets, less than 500,000, in summary. We conducted some
experiments with more MCMC samples than those values, but there was no significant improvement w.r.t. the
estimation accuracy. Therefore, the results described in the following refer to
the minimum number of samples suggested by the Raftery-Lewis test.

Last, recall from Section~\ref{sec:approx} that $\btheta_0$ can be any distribution.
However, the closer it is to $\btheta$, the better is the convergence of the MCMC.
When estimating the offspring distribution in medium size problems, we will
assume that $\btheta_0$ is binomial and set its parameters so that the average
is $\bar{d}$. This implies assuming that the average number of offspring can be
estimated, but in fact a rough estimate can be obtained by simply taking the average of the observed node degrees from the first generations in the
sample, whose edges have a relatively high probability of being sampled.
For small sized tree, we simply set $\btheta_0$ to be uniform.

\subsection{Results}

The estimation results span over a number of dimensions equal to the
number of parameters assumed in the multinomial distribution. We use the
Kullback-Leibler (KL) divergence
as an objective criterion to compare the estimation methods in a single
dimension.

Let the estimated offspring distribution be
$\hat{\btheta}=(\hat\theta_1,\dots,\hat\theta_W)$.
The KL-divergence of $\hat{\btheta}$ from $\btheta$ is defined by
\beq
D_{\textrm{KL}}(\btheta || \hat{\btheta}) = \sum_{i=1}^W (\log \theta_i -\log
        \hat\theta_i) \theta_i,
\eeq
when $\hat\theta_i > 0,\,i=1,\dots,W$. When this condition does not always hold,
as in our case, absolute discounting is frequently used to smooth $\hat\btheta$.
Hence, we distribute $\epsilon=10^{-7}$ of probability mass among the zero estimates, discounting
this value equally from the non-zero estimates.

Table~\ref{tab:kldiv} shows the median KL-divergence obtained for each set of problems (indexed by
$\btheta^{(i)},\,i=1,\dots,3$), for \Exact\ and \Approx, when the
sampling probability $p$ is medium.
\begin{table}[ht]
\begin{center}
\begin{tabular}{rrrr}
\hline
& $\btheta^{(1)}$& $\btheta^{(2)}$& $\btheta^{(3)}$  \\ 
\hline
\Exact & 1.86 & -  & -  \\ 
\Approx & 2.98 & 0.58 & 0.78  \\ 
\hline
\end{tabular}
\end{center}
\caption{Median KL-divergence of estimators}
\label{tab:kldiv}
\end{table}
Dashes indicate that \Exact\ could not find
estimates for medium size problems in a reasonable
amount of time. However, it outperfomed \Approx\ in the estimation of
$\btheta^{(1)}$.
Note that although KL-divergence implies some ordering
within each column in terms of accuracy, neither the relative ratios have a direct
interpretation, nor values accross different columns can be compared. We will
next evaluate the results w.r.t.\ the MSE of each parameter estimate, which will
allow us to conclude that the performance of \Approx\ is in fact very close to
the one of \Exact\ for small datasets.

\subsection*{The effect of sampling probabilities}

As we increase $p$, we gather more information about the original graph and
hence estimators will clearly perform better. We study the performance gains
w.r.t.\
the MSE of the parameter estimates.

\begin{figure}[!t]
\vspace{-1.0cm}
\begin{center}
\subfloat[\Exact]{\includegraphics[width=1.7in]{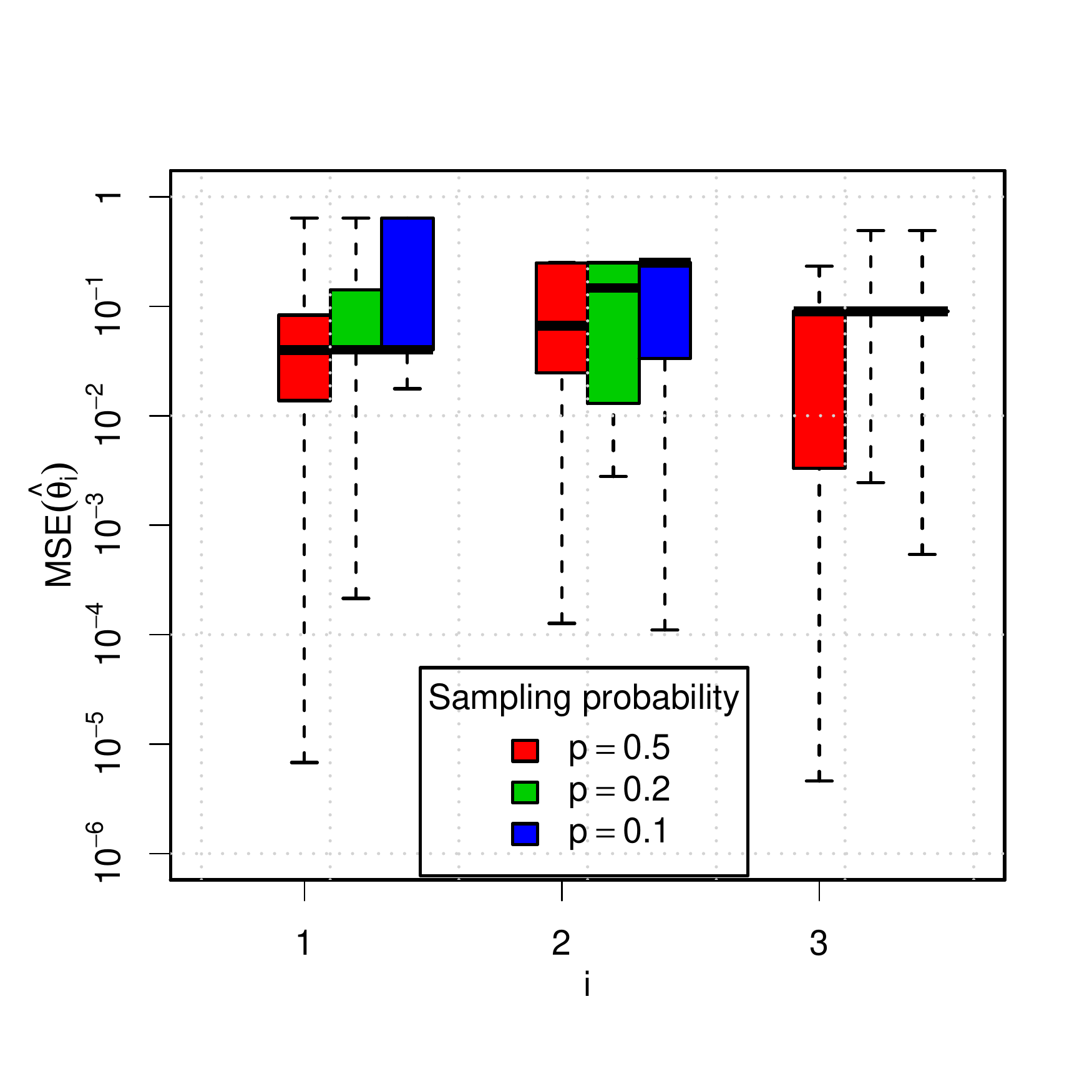}\label{fig:theta1-exact}}
\subfloat[\Approx]{\includegraphics[width=1.7in]{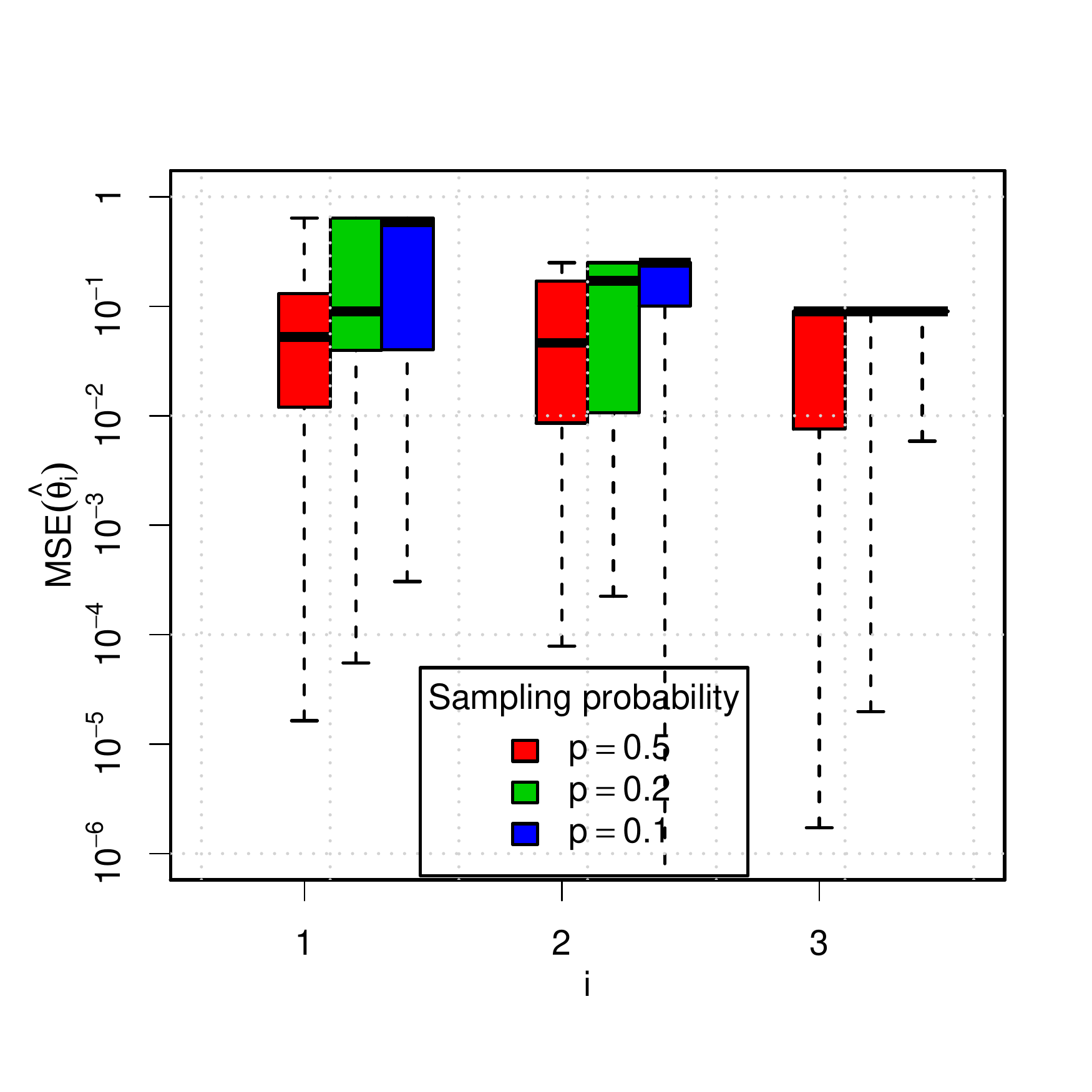}\label{fig:theta1-approx}}
\end{center}
\vspace{-0.5cm}
\caption{Boxplots of the MSE per parameter for $\btheta^{(1)}$.}
\label{fig:theta1}
\end{figure}

Figs.~\ref{fig:theta1}(a-b) show boxplots of the MSE of the estimates
$\hat{\theta_i},\,i=1,\dots,W$ obtained by
\Exact\ and \Approx, respectively, for datasets coming from $\btheta^{(1)}$. Each boxplot shows minimum,
1st quartile, median, 3rd quartile and maximum values, computed over 100 estimates (10 samples for each
of the 10 trees). Colors correspond to different sampling
probabilities. In both cases, the median MSE increases as we
decrease $p$, as expected.

\begin{figure}[!t]
\vspace{-1.0cm}
\begin{center}
\includegraphics[width=2.7in]{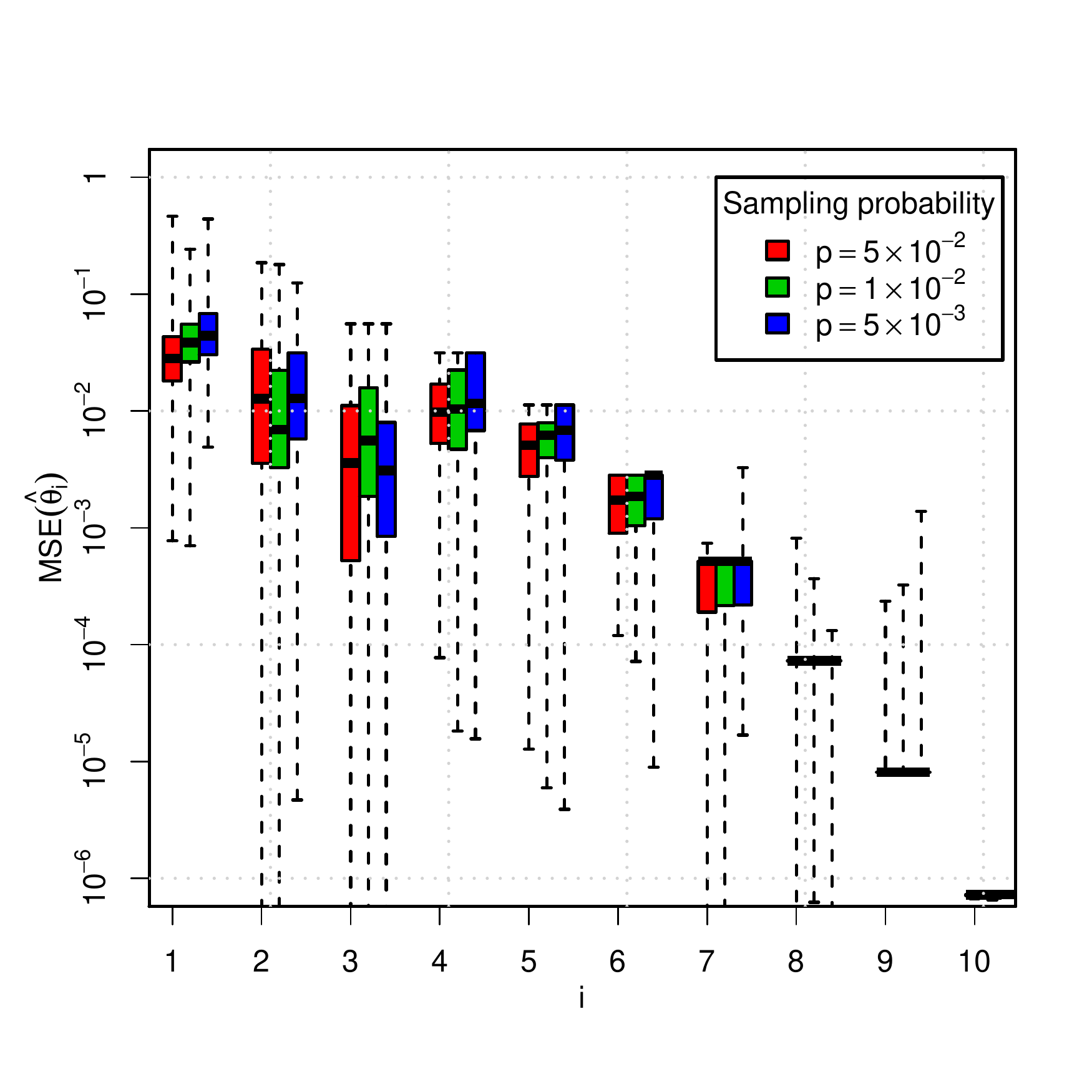}\label{fig:theta2-approx}
\end{center}
\vspace{-1.0cm}\caption{Boxplots of the MSE of \Approx\ for $\btheta^{(2)}$.}
\label{fig:theta2-bp}
\end{figure}

Similarly, Fig.~\ref{fig:theta2-bp} shows the results
obtained by \Approx\ for datasets that come from $\btheta^{(2)}$. In general,
increasing the sampling probability reduces the MSE, but not by a significant
amount. Results for $\btheta^{(3)}$ are similar and will be ommitted.

We conjecture that most of the information that allows us to estimate
$\btheta$ comes from the top levels of the tree. As we increase $p$,
we obtain many more observations from the bottom levels of the tree, but only a
few new observations from the top levels. While edges closer to the root are
observed with higher probability, edges from lower levels are more rarely
sampled and there is much more uncertainty in those samples. This implies that increasing $p$
should not improve the estimates significantly after a certain point.

This short digression might lead the reader wonder whether the values of $p$
we use would sample so many edges from the top levels that would be
enough to take the empirical distribution of the observed degrees
at those levels as an estimate for $\btheta$.
Hence, we compare the MSE results for \Approx\ with
the empirical distribution of the observed degrees from the
top 1, 2 and 3 levels in a cumulative
fashion. Intuitively, the empirical
distribution is biased towards smaller degrees, especially if lower levels are
taken into account, this being the reason why we stop at 3 levels.

\begin{figure}[!t]
\vspace{-0.5cm}
\begin{center}
\includegraphics[width=2.7in]{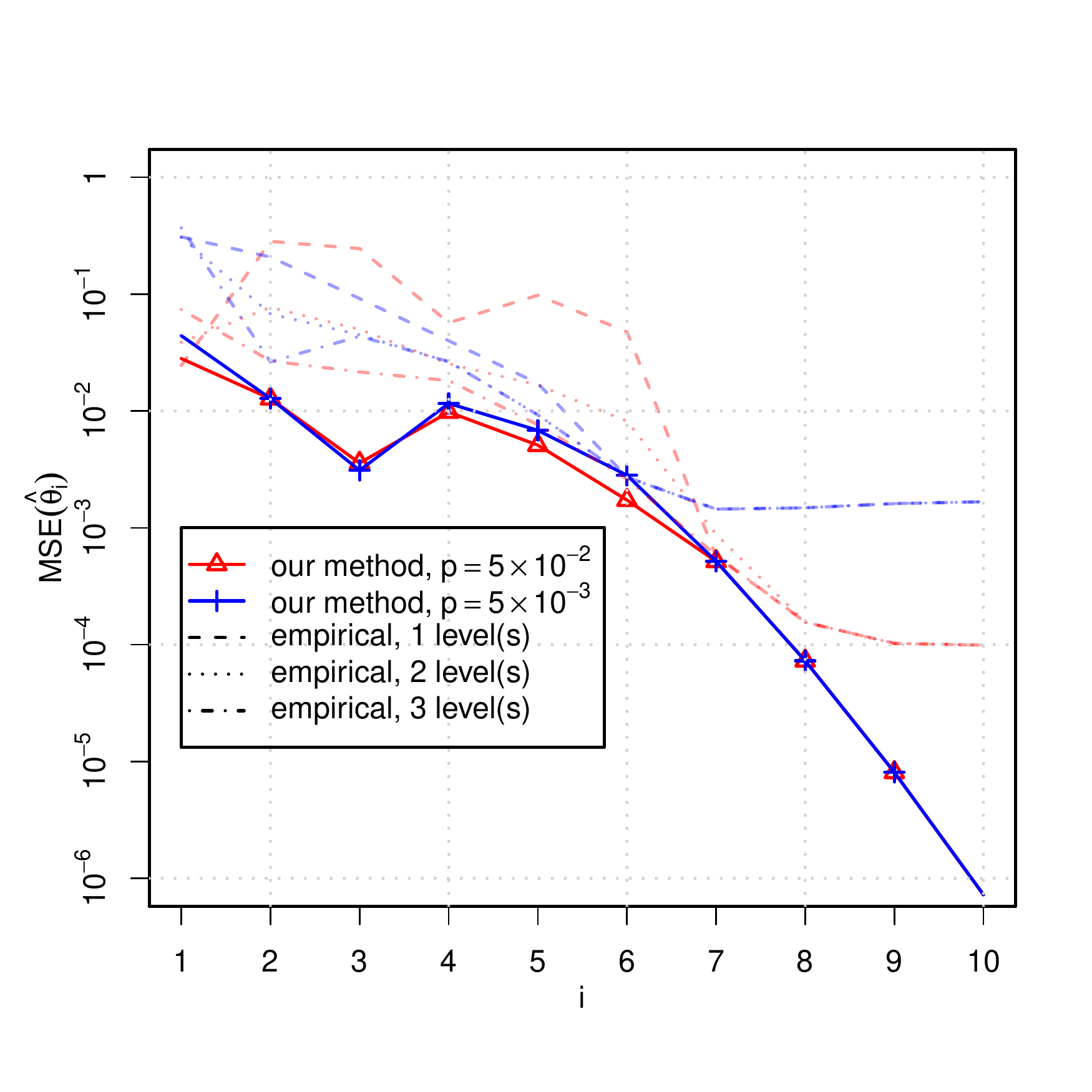}
\end{center}
\vspace{-1.0cm}\caption{Median MSE of \Approx\ and empirical estimates for $\btheta^{(2)}$.}
\label{fig:theta2-med}
\end{figure}

Fig.~\ref{fig:theta2-med} shows the median values of the MSE (also seen in
the previous figure), but only for ``small'' and ``large'' $p$ values, for the
sake of clarity. In addition, dashed lines display the median MSE obtained when the
empirical distributions are used as estimators. 
Estimates for $p=5\times10^{-3}$ exhibit a
one-order magnitude gain in accuracy (for most parameters) relative to the
best empirical estimate, but estimates for $p=5\times10^{-2}$ only yield
significant improvements at the tail of the distribution. In general,
empirical distributions are not good estimates, especially for distribution
tails due to its bias towards small degrees. One exception we found was in the
case of $\btheta^{(3)}$, where the probability mass at the tail
is so large that high degree nodes are likely to be observed at
the top levels. However, we observed in additional experiments that this is not the case for long
tailed distributions with larger support, such as $W=100$.




%% file: conclusions.tex
\section{Conclusions}\label{sec:conclusions}

In this paper we propose and analyze two methods to estimate the offspring distribution of
a branching process from a sample of random observed paths to the root. The former, based on
exact inference, is limited to small problems since the number of terms to be
computed in the likelihood function grows exponentially with the maximum degree
and number of levels. The latter, approximates the likelihood function using
MCMC samples, and was able to handle both small and medium size problems. For
small problems, its performance was similar to that of exact inference.

%% file: branching.bbl
\begin{thebibliography}{1}

\bibitem{Achlioptas09}
D.~Achlioptas, A.~Clauset, D.~Kempe, and C.~Moore.
\newblock On the bias of traceroute sampling: Or, power-law degree
  distributions in regular graphs.
\newblock {\em J. ACM}, 56(4):21:1--21:28, July 2009.

\bibitem{Beerli99}
P.~Beerli and J.~Felsenstein.
\newblock {Maximum-Likelihood Estimation of Migration Rates and Effective
  Population Numbers in Two Populations Using a Coalescent Approach}.
\newblock {\em Genetics}, 152(2):763--773, June 1999.

\bibitem{Chierichetti11}
F.~Chierichetti, J.~Kleinberg, and D.~Liben-Nowell.
\newblock Reconstructing patterns of information diffusion from incomplete
  observations.
\newblock In {\em NIPS'11}, pages 792--800, 2011.

\bibitem{Lakhina03}
A.~Lakhina, J.W. Byers, M.~Crovella, and P.~Xie.
\newblock Sampling biases in ip topology measurements.
\newblock In {\em INFOCOM 2003}, volume~1, pages 332 -- 341 vol.1, march-3
  april 2003.

\bibitem{neves2006}
A.~Neves and C.~Moreira.
\newblock Applications of the galton-watson process to human dna evolution and
  demography.
\newblock {\em Physica A}, 368(1):132 -- 146, 2006.

\bibitem{vespignani}
R.~Pastor-Satorras and A.~Vespignani.
\newblock {\em Evolution and Structure of the Internet: A Statistical Physics
  Approach}.
\newblock Cambridge University Press, New York, NY, USA, 2004.

\bibitem{raftery95}
A.~Raftery and S.~Lewis.
\newblock {The number of iterations, convergence diagnostics and generic
  Metropolis algorithms}.
\newblock In {\em In Practical Markov Chain Monte Carlo (W.R. Gilks, D.J.
  Spiegelhalter and S. Richardson, eds.)}, pages 115--130, 1995.

\bibitem{veciana2004}
X.~Yang and G.~de~Veciana.
\newblock Service capacity of peer to peer networks.
\newblock In {\em INFOCOM}, pages 2242--2252 vol.4, march 2004.

\end{thebibliography}
